\documentclass[11pt,epsf]{article}

\usepackage{epsfig}
\usepackage{color}

\begin{document}

\baselineskip 6mm
\renewcommand{\thefootnote}{\fnsymbol{footnote}}


\newcommand{\nc}{\newcommand}
\newcommand{\rnc}{\renewcommand}



\newcommand{\tcb}{\textcolor{blue}}
\newcommand{\tcr}{\textcolor{red}}
\newcommand{\tcg}{\textcolor{green}}


\def\be{\begin{equation}}
\def\ee{\end{equation}}
\def\ba{\begin{array}}
\def\ea{\end{array}}
\def\bea{\begin{eqnarray}}
\def\eea{\end{eqnarray}}
\def\nn{\nonumber\\}


\def\ct{\cite}
\def\la{\label}
\def\eq#1{(\ref{#1})}
\def\text#1{{\rm #1}}


\def\a{\alpha}
\def\b{\beta}
\def\g{\gamma}
\def\G{\Gamma}
\def\d{\delta}
\def\D{\Delta}
\def\e{\epsilon}
\def\et{\eta}
\def\ph{\phi}
\def\Ph{\Phi}
\def\ps{\psi}
\def\Ps{\Psi}
\def\k{\kappa}
\def\l{\lambda}
\def\L{\Lambda}
\def\m{\mu}
\def\n{\nu}
\def\th{\theta}
\def\Th{\Theta}
\def\r{\rho}
\def\s{\sigma}
\def\S{\Sigma}
\def\ta{\tau}
\def\o{\omega}
\def\O{\Omega}
\def\pr{\prime}
\def\z{\zeta}


\def\half{\frac{1}{2}}

\def\goto{\rightarrow}

\def\na{\nabla}
\def\grad{\nabla}
\def\curl{\nabla\times}
\def\div{\nabla\cdot}
\def\pa{\partial}

\def\bra{\left\langle}
\def\ket{\right\rangle}
\def\lb{\left[}
\def\lc{\left\{}
\def\ls{\left(}
\def\lp{\left.}
\def\rp{\right.}
\def\rb{\right]}
\def\rc{\right\}}
\def\rs{\right)}
\def\fr{\frac}

\def\vac#1{\mid #1 \rangle}


\def\td#1{\tilde{#1}}
\def\check{ \maltese {\bf Check!}}


\def\Tr{{\rm Tr}\,}
\def\det{{\rm det}}


\def\bc#1{\nnindent {\bf $\bullet$ #1} \\ }
\def\ch {$<Check!>$ }
\def\ss {\vspace{1.5cm}}

\begin{titlepage}

\hfill\parbox{5cm} { }

\vspace{25mm}

\begin{center}
{\Large \bf Notes on the holographic Lifshitz theory}

\vskip 1. cm
  {Chanyong Park$ $\footnote{e-mail : cyong21@ewha.ac.kr}}

\vskip 0.5cm

{\it Institute for the Early Universe, Ewha womans University, DaeHyun 11-1, Seoul 120-750, Korea}

\end{center}

\thispagestyle{empty}

\vskip2cm


\centerline{\bf ABSTRACT} \vskip 4mm

\vspace{1cm}
On the Lifshitz black brane geometry of an Einstein-Maxwell-dilaton gravity,
we holographically investigate electric DC conductivities and the role of impurity
in a non-relativistic Lifshitz medium with two different charge carriers, impurity and  Lifshitz matter.
The conductivity carried by Lifshitz matter is proportional to the square of temperature, 
while that carried by impurity crucially depends on the bulk coupling parameter $\g$. 
For $\g < -2$, impurity at high temperature can change the electric property of the Lifshitz medium significantly so that the Lifshitz matter with impurity can show a totally different 
electric property from the pure Lifshitz matter.

\vspace{2cm}


\end{titlepage}

\renewcommand{\thefootnote}{\arabic{footnote}}
\setcounter{footnote}{0}


\section{Introduction}

The AdS/CFT correspondence is a very useful and fascinating tool for understanding 
the strongly interacting system \cite{Maldacena:1997re}. 
In the last decade, it has been widely used in studying some universal properties 
of Quantum Chromodynamics (QCD) and condensed matter systems in the strong coupling 
regime \cite{Sakai:2004cn,Hartnoll:2008vx,Horowitz:2006ct,Hartnoll:2009sz,Sachdev:2010ch,Sachdev:2008ba}. 
The asymptotic $AdS$ geometry plays an important role in such investigations because its
dual theory is described by the conformal symmetry.
Can we generalize the  AdS/CFT correspondence to the non-AdS geometry? 
It is an interesting and also important question in understanding the non-conformal or 
non-relativistic condensed matter systems 
through the holographic methods \cite{Koroteev:2007yp}-\cite{Gouteraux:2011qh}. 
In this paper, we will study the electric conductivities in the non-relativistic Lifshitz theory 
with two kinds of charge carriers.

Following the gauge/gravity duality it was shown that the Einstein-dilaton theory
with a Liouville potential corresponds to a relativistic non-conformal theory 
\cite{Kulkarni:2012re,Kulkarni:2012in}. In addition,
it was also found that the DC conductivity of the dual system can show different behaviors
depending on what kind of vector fluctuation is turned on. If a vector fluctuation is not
coupled to dilaton, the corresponding DC conductivity in a $2+1$-dimensional 
relativistic non-conformal theory is temperature independent, while it can have a nontrivial 
temperature dependence for the vector fluctuation coupled to dilaton.  These facts were also 
checked by using the membrane paradigm \cite{Park:2012cu}. 
In the similar setup without a Liouville 
potential, the exact gravity solution has been known as the Lifshitz geometry \cite{Kachru:2008yh,Taylor:2008tg}. 
Although the Lifshitz geometry has different scaling in the temporal and spatial coordinates 
the generalized scaling symmetry, the so called hyperscaling symmetry, is still preserved.
Due to such a nontrivial scaling, it has been believed that the Lifshitz geometry
is dual to a Lifshitz field theory.
In particular, when the dynamical exponent is $z=2$, the
corresponding dual theory becomes non-relativistic.  
In the holographic QCD models for a strongly interacting quark-gluon plasma
\cite{Maldacena:1998im,Rey:1998ik,Lee:2009bya,Ge:2008ak,Gubser:2006bz,Kim:2008ax},
the asymptotic AdS space has been used as the dual geometry. These works were further
generalized to the charged AdS geometry, for example, the thermal charged AdS and
Reissner-Nordstr\"{o}m black brane geometry \cite{Lee:2009bya}. In these cases,
a bulk vector field was identified with matter of the dual theory. Similarly, the bulk vector field
of the Lifshitz geometry might be regraded as matter of the Lifshitz field theory. 
Here, we will simply call it Lifshitz matter or Lifshitz medium. 
In this non-relativistic Lifshitz medium, the
binding energy and drag force of external quarks were investigated in \cite{Fadafan:2009an}.  

There were many studies on the DC conductivity and superconductivity of the non-relativistic 
Lifshitz medium without a dilaton coupling \cite{Mann:2009yx}-\cite{Ge:2010aa}. 
We further investigate the DC conductivities of the Lifshitz medium 
with a nontrivial dilaton coupling, which provides more information for the charge carrier.
In this paper, a new vector fluctuation is turned on in the Lifshitz black brane geometry 
to describe impurity in the non-relativistic Lifshitz medium. This new vector fluctuation can
have a different dilaton coupling from the background gauge field.
If we parameterize the different dilaton coupling with $\g$, there exists a discrepancy
between the results of the membrane paradigm and the Kubo formula for $\g \ge 1$.
This is due to the change of the asymptotic boundary condition.
While the membrane paradigm does not care about the change of the asymptotic boundary condition  
\cite{Kovtun:2003wp,Iqbal:2008by}, the
Kubo formula crucially depends on it \cite{Son:2002sd,Policastro:2002se,Policastro:2002tn}. 
On the other hand, the fluctuation of the background gauge field corresponds to
the Lifshitz matter. The DC conductivity carried by it shows totally different
behavior from that of impurity because it is coupled to the metric fluctuation
through the background gauge field even at quadratic order. The Kubo formula
says that the DC conductivity carried by the Lifshitz matter is proportional to
the square of temperature.
In certain condensed matter systems like semiconductor, 
impurity is important to explain their electric property. Therefore it is interesting
to understand the role of impurity in the medium. We find that 
at high temperature impurity with $\g<-2$ can change the
electric property of the non-relativistic Lifshitz medium significantly.

The rest of the paper is organized as follows: In Sec. 2, we represent 
the Lifshitz black brane solution including the manifest hyperscaling symmetry and its 
thermodynamics with explaining our conventions. 
In Sec. 3, the DC conductivities carried two different charge carriers in the non-relativistic Lifshitz medium are studied. The results show that impurity with $\g<-2$
significantly changes the electric property of the Lifshitz medium at high temperature.
Finally, we finish this work with some concluding remarks.


\section{Thermodynamic properties}

There exist many scale-invariant field theories without the Lorentz invariance near the critical points \cite{Hartnoll:2009sz,Sachdev:2008ba}. One of such examples is the Lifshitz theory 
\be 
S[\chi] = \int d^{3} x \lb \ls \pa_t  \chi \rs^2 - K \ls \nabla^2  \chi \rs^2 \rb ,
\ee
which describes a fixed line parameterized by $K$ with a dynamical exponent $z=2$ \cite{Balasubramanian:2009rx}. Following the gauge/gravity
duality, such a non-relativistic theory can be mapped to a Lifshitz geometry 
as a dual gravity. 
There are several bottom-up models, gravity with higher form fields \cite{Kachru:2008yh} and gravity
with a massive gauge field and non-dynamical scalar field. These models have been widely investigated 
by many authors \cite{Balasubramanian:2009rx,Bertoldi:2009vn,Korovin:2013bua}.
Another example appears as a geometric solution of the Einstein-Maxwell-dilaton theory. 
In this paper we will concentrate on the latter case.

Our starting action is the Einstein-Maxwell-dilaton theory with a negative cosmological 
constant $\L$ 
\be			\la{act:original}
S_{EMd} = \frac{1}{16 \pi G} \int d^{D} x \sqrt{-g} \ls  R - 2 \L
- \half \pa_{\m} \ph \ \pa^{\m} \ph - \frac{1}{4} e^{\l \ph} F_{\m\n} F^{\m\n}  \rs ,
\ee
with
\be
F_{\m\n} = \pa_{\m} A_{\n} - \pa_{\n} A_{\m}  ,
\ee
where $\l$ is a constant describing the coupling between the gauge field and dilaton. 
From this action, the black brane geometry satisfying all equations of motion is given by  
\cite{Taylor:2008tg,Pang:2009ad,Pang:2009wa,Tarrio:2011de}
\bea			\la{sol:lifshitz}
ds^2 &=& - r^{2z} f(r) dt^2 + \frac{dr^2}{r^2 f(r)} + r^2 (dx^2 + dy^2 )  , \nn
\ph (r) &=& - \fr{4}{\l} \log r , \nn
F_{rt} &=& \pa_r A_t = q \ r^{z+1} ,
\eea
with
\bea
f(r) &=& 1 - \frac{r_h^{z+2}}{r^{z+2}} , \la{met:blacklif} \\
\l &=& \frac{2}{\sqrt{z-1}} , \nn
q &=& \sqrt{2 (z-1) (z+2)} ,\nn
\L &=& - \frac{(z+1)(z+2)}{2} ,
\eea
where $r_h$ implies the black brane horizon.
It is worthwhile noting that the Lifshitz black brane is not a charged but uncharged 
solution because the charge $q$
is not a free parameter describing a hair of the black brane. In other words,
once the intrinsic parameters of the theory, $\L$ and $\l$, are given
the dynamical exponent $z$ and the charge $q$ are automatically determined. 
In that sense, the Lifshitz black brane geometry might be corresponding to the
microcanonical ensemble while the charged black branes are described
by a grandcanonical or canonical ensemble.
For $z=1$, dilaton and bulk gauge field automatically vanish and the Lifshitz geometry 
simply reduces to an ordinary AdS geometry where the conformal symmetry is restored. 

Note that the above Einstein-Maxwell-dilaton theory preserves a scaling symmetry.
When $r$ scales as $\O r$, other variables should scale like
\bea		\la{res:hyperscal}
&& r_h \to \O r_h , \quad t \to \O^{-z} t , \quad x \to \O^{-1} x ,  \quad y \to \O^{-1} y ,\nn
&& e^{\ph} \to \O^{-4/\l} \ e^{\ph} , \quad q\to \O^{0} q , \quad F_{rt} \to \O^{z+1} F_{rt}  , \quad A_t \to \O^{z+2} A_t .
\eea
These scaling behaviors are different from those of the different Lifshitz models.
Then, the time component gauge field satisfying the above scaling is given by
\be		\la{sol:backvec}
A_t = \fr{q}{z+2 } \ls r^{z+2} - r_h^{z+2} \rs ,
\ee
where the last term corresponds to a new integration constant and we choose a specific
value such that the norm of $A_t$ is regular even at the black brane horizon.
The energy and temperature of this system should scale as the inverse of time,
$E \to \O^z E$ and $T \to \O^{z} T$.
After expanding the metric near the horizon
and requiring that there is no conical singularity,
the Hawking temperature is determined to be
\be			\la{res:hawkingtemp}
T = \frac{z+2}{4 \pi} r_h^{z} ,
\ee
which shows the correct scaling behavior mentioned previously.
The Bekenstein-Hawking entropy is
\be
S = \frac{V_2}{4 G} r_h^2 ,
\ee
where $V_2$ implies a spatial volume of the boundary space and scales like 
$V_2 \to \O^{-2} V_2$. Therefore, the Bekenstein-Hawking entropy is invariant under
the scaling transformation.

Using the first law of thermodynamics together with the above Hawking temperature and 
Bekenstein-Hawking entropy, 
the internal energy $E$ and the free energy $F$ are given by
\bea		\la{res:thenergy}
E &=& \fr{V_2}{8 \pi G} r_h^{z+2} , \nn
F &=& - \fr{z V_2}{16 \pi G} r_h^{z+2} .
\eea
Using the definition of pressure $P = - \pa F/\pa V_2$, we can easily evaluate the equation of state parameter
of the Lifshitz black brane
\be			\la{res:eqofstate}
w = \fr{P V_2}{ E} =  \fr{z}{2} .
\ee
This implies, according to the gauge/gravity duality, that the dual theory is not conformal 
except the AdS case with the dynamical exponent $z=1$. Note that
it was shown, in the gravity theory with a massive gauge field and  non-dynamical scalar field\footnote{In \cite{Balasubramanian:2009rx}, only the $z=2$ case has been considered and the black brane factor is given by $f(r) = 1 - \frac{r_h^{2}}{r^{2}} $ in our notations, which
is different from the present one, $f(r) = 1 - \frac{r_h^{4}}{r^{4}}$ in \eq{met:blacklif}.
Although the asymptotic geometries of these two different theories are exactly same, 
the Hawking temperature due to the difference of inside geometry 
can have different values. 
For example, $T =\frac{r_h^2}{2 \pi}$ in \cite{Balasubramanian:2009rx} 
and $T=\frac{r_h^2}{\pi}$ in our case.},
that the equation of state parameter is given by $1$ for $z=2$. 
The specific heat of this system becomes in terms of temperature
\be
C_v = \fr{V_2}{2 z G} \ls \fr{4 \pi}{z+2} \rs^{\fr{2}{z}}  T^{\fr{2}{z}} .
\ee
Since it is positive for $z > 0$, the dual Lifshitz theory is always thermodynamically stable.
In the zero temperature limit $r_h \to 0$, the internal and free energies in \eq{res:thenergy}
become zero. Comparing them with the results at finite temperature, since the free energy at finite temperature is always negative for $z>0$, the Lifshitz black brane is always preferable. This fact implies that there is no Hawking-Page transition.
A similar situation also occurs in the relativistic non-conformal theory \cite{Kulkarni:2012in}.

\section{DC conductivities in the non-relativistic Lifshitz medium}

In order to understand macroscopic properties of a non-relativistic Lifshitz medium, 
it is useful to investigate the holographic linear response of various fluctuations. 
Here, we will concentrate on the electric properties of a non-relativistic Lifshitz theory 
with two different charge carriers. To do so, we consider a more general 
Einstein-Maxwell-dilaton theory
\be
S_{g} = \frac{1}{16 \pi G} \int d^{D} x \sqrt{-g} \ls  R - 2 \L
- \half \pa_{\m} \ph \ \pa^{\m} \ph - \frac{1}{4} e^{\l \ph} F_{\m\n} F^{\m\n}  
- \fr{1 }{4} e^{\g \ph}  H_{\m\n} H^{\m\n} \rs ,
\ee
with
\bea
F_{\m\n} &=& \pa_{\m} A_{\n}-\pa_{\n} A_{\m} , \nn
H_{\m\n} &=&\pa_{\m} B_{\n}-\pa_{\n} B_{\m} ,
\eea
where $A_{\m}$ and $B_{\m}$ are two different $U(1)$ vector fields
with different dilaton couplings.
In this model, the previous Lifshitz black brane geometry appears as a specific solution for
$B_{\m}=0$. In the dual Lifshitz theory point of view, these two different vector fields
corresponds to two different matters.

Now, let us turn on vector and metric fluctuations on the Lifshitz black brane geometry.
If we denote $a_{\m}$ and $b_{\m}$ as fluctuations of $A_{\m}$ and $B_{\m}$ respectively,
they are governed by the following action at quadratic order
\be
S_{fluc} = S_a + S_b , 
\ee
with
\bea  
S_{a} &=& \frac{1}{16 \pi G} \int d^{4} x \sqrt{-g} \ls  {\cal R} - \frac{e^{\l \ph} }{4} 
f_{\m\n} f^{\m\n} \rs , \la{act:flucta} \\
S_{b} &=& - \fr{1}{16 \pi G} \int d^4 x \sqrt{-g}  \  \fr{e^{\g \ph} }{4}  h_{\m\n} h^{\m\n}  ,
\la{act:fluctb}
\eea  
where $f_{\m\n} = \pa_{\m} a_{\n} - \pa_{\n} a_{\m}$ 
and $h_{\m\n} = \pa_{\m} b_{\n} - \pa_{\n} b_{\m}$. There is no
mixing term between $a_{\m}$ and $b_{\m}$ at quadratic order, so one can describe those two fluctuations independently.
Note that since the fluctuation $a_{\m}$, which is called Lifshitz matter, is coupled to 
the metric fluctuations through the background gauge field  even at quadratic order, 
one should take into account the metric fluctuations simultaneously. As will shown, the coupling to
the metric fluctuations dramatically changes the DC conductivity carried by the Lifshitz matter.
The other fluctuation, $b_{\m}$, is a new one which corresponds to impurity in the dual Lifshitz field theory.
Since impurity is nothing to do with the background gauge field at quadratic order, there is no mixing 
with the metric fluctuation \cite{Pang:2009wa,Lee:2010qs}.  In addition, the coupling parameter $\g$
can have an arbitrary number, which may depend on the kind of impurity.   
From now on we concentrate on the $z=2$ case, which provides an interesting example
for a non-relativistic Lifshitz theory, and take the zero momentum limit because the DC conductivity is well defined even in this limit. In the following sections, we will investigate 
the DC conductivities carried by impurity and  Lifshitz matter with the Kubo formula 
and show that the different charge carriers lead to the different electric properties.

\subsection{DC conductivity carried by impurity}

First, let us study the DC conductivity carried by impurity with the Kubo formula.
From the action \eq{act:fluctb} for impurity, the transverse mode, $b_i$ ($i = x$ or $y$), is governed by
\be
0 = \pa_{\m} \lb \sqrt{-g} e^{\g \ph} g^{\m \r} g^{i \s} \ls 
\pa_{\r} b_{\s} - \pa_{\s} b_{\r}\rs \rb .
\ee
For $z=2$ and in the zero momentum limit, under the following Fourier mode expansion 
\be
b_i (t,r) = \int \fr{d \o}{2 \pi} e^{- i \o t} \ b_i (\o,r) ,
\ee
the governing equation simply reduces to
\be		\la{eq:impurityeq}
0 =  {b_i}'' + \ls \fr{3- 2 \g}{r} + \fr{f'}{f} \rs b_i' + \fr{\o^2}{r^6 f^2} b_i .
\ee

At the horizon, $b_i$ has two independent solutions
\be 		\la{sol:nearhsol}
b_i  (r) = c_1 f^{ \pm \n} ,
\ee
with $\n = i \fr{\o}{4 r_h^2}$, where $c_1$ is an appropriate normalization constant and 
the minus or plus sign satisfies the incoming or outgoing
boundary condition at the horizon. After choosing an incoming solution,  
the solution of \eq{eq:impurityeq} in the hydrodynamic limit ($\o \ll T$) can be perturbatively expanded to
\be		\la{sol:psolb}
b_i (r) =  f^{ - \n} \lb G_0 (r) + \o G_1 (r)    \rb  + {\cal O} (\o^2) .
\ee
In this hydrodynamic expansion, $G_0 (r)$, $G_1 (r)$ and all higher order terms should be regular at the horizon which is called a regularity condition. 
In addition, the above perturbative solution should be reduced to \eq{sol:nearhsol} at the horizon, so $G_0 (r)$ should be a normalization constant $c_1$ at the horizon
and at the same time the other terms, $G_1 (r_h)$ and higher order terms, should vanish.
We call such a constraint a vanishing condition.
Using these two conditions, the perturbative solutions can be exactly determined
up to one integration constant
\bea
G_0 (r) &=& c_1 , \la{res:udicon} \\
G_1 (r) &=& c_3  - \fr{i c_1  \lb 4 \log r - \log (r^4 - r_h^4) \rb }{4 r_h^2}  \nn
&& - \fr{c_4  \lb {}_2 F_1 \ls 1 + \g, 1, 2 + \g, - \fr{r^2}{r_h^2} \rs  
+  {}_2 F_1 \ls 1 + \g, 1, 2 + \g,  \fr{r^2}{r_h^2} \rs \rb r^{2 + 2 \g}}{4 (1 + \g)  r_h^4} ,
\eea
with
\bea
c_3 &=& \fr{i c_1 \lb - PG \ls 0, 1 + \fr{\g}{2} \rs + 
   PG \ls 0, \fr{1 + \g}{2} \rs + 2 \lc EG - \log 2 + PG \ls 0, 1 + \g \rs \rc \rb }{8 r_h^2} , \\
c_4 &=& - i c_1 r_h^{-2 \g}  ,
\eea
where $PG$ and $EG$ mean the poly gamma and Euler gamma function respectively.
In order to determine the remaining integration constant $c_1$, we should impose another 
boundary  condition. At the asymptotic boundary, the vector fluctuation $b_i$ has 
the following asymptotic expansion
\be			\la{res:asymtsoldu1}
b_i (r) = b_1  + b_2 \ r^{2 \g-2} ,
\ee
where $b_1$ (or $b_2$) is a constant determined by the asymptotic boundary condition.

\subsubsection{For $\g < 1$}

If $\g$ is smaller than $1$, the asymptotic behavior of $b_i (r)$ is determined by 
the first term $b_1$. According to the usual gauge/gravity 
duality, the first coefficient corresponds to the source while the second
describes the vacuum expectation value (vev) of the dual operator. In this case, it is natural
to impose the Dirichlet boundary condition like
\be			\la{bc:diffu1so}   \,
b_0  \equiv  \lim_{r_0 \to \infty}  b_i  (r_0) ,
\ee
where $r_0$ implies an appropriate UV cutoff of the dual theory and
$b_0$ corresponds to the boundary value of $b_i$ which is equal to $b_1$ 
for $\g < 1$.
Comparing the asymptotic expansion of the perturbative solution \eq{sol:psolb} with 
the above boundary condition \eq{bc:diffu1so}, 
the remaining integration constant $c_1$ can be rewritten 
in terms of the boundary value $b_0$ as
\be			\la{bval:imp}
c_1 = \fr{8 i r_h^{2} \ b_0}{8 i r_h^2 +\o \lb HN \ls \fr{\g}{2} \rs - HN \ls \fr{ -1 + \g }{2} \rs  - 
 2 HN \ls \g \rs  + 2 \log 2 + 2 \pi \tan \ls \fr{\pi \g}{2} \rs \rb } ,
\ee
where $HN$ means a harmonic number.

The boundary action corresponding to the on-shell action of \eq{act:fluctb} is given by
\be
S_{B} = - \frac{1}{16 \pi G} \int_{r = r_0} d^3 x \ \sqrt{-g}  \ e^{\g \ph} g^{rr} g^{ii}
b_{i}  {b_i}'  \approx  - \frac{1}{16 \pi G} \int d^3 x \ r_h^{- 2 \g } \
r_0^{3 - 2 \g}   \ b_{0}  \ {b_i}'  .
\ee
This result shows that 
the finite contributions to the boundary action can come from ${b_i}'  \sim r_0^{-3+2 \g}$
when $r_0 \to \infty$.
Since the asymptotic expansion of ${b_i}'$ from \eq{sol:psolb} has
\be
{b_i}' = - \fr{i c_1 \o}{r_h^{2 \g}}  \fr{1}{r_0^{3-2 \g}} + {\cal O} \ls \fr{1}{r^5} \rs ,
\ee
the current-current retarded Green function \cite{Policastro:2002se,Policastro:2002tn} results in
\be
\bra J^i J^i \ket = \fr{i \o}{16 \pi G} \fr{1}{r_h^{2 \g}} + {\cal O} (\o^2) ,
\ee
where \eq{bval:imp} is used.
Finally, the DC conductivity from the Kubo formula reads
\be			\la{res:kubob}
\s_{DC} \equiv \lim_{\o \to 0} \fr{\bra J^i J^i \ket }{i \o} = \fr{1}{16 \pi^{\g+1} G} 
\ \fr{1}{T^{\g}} .
\ee

\subsubsection{For $ \g \ge 1$}

Let us take into account the case with $\g \ge 1$.  In this case, the DC conductivity carried by impurity
shows a totally different behavior compared with the previous case because
the interpretation of the asymptotic solution should be modified.
From now on, we will concentrate on the case with $\g =2$ for later comparison 
with the DC conductivity carried by Lifshitz matter. 

Similar to \eq{sol:psolb}, the perturbative expansion of solution in 
the zero momentum limit is given by 
\be
b_i (r) =  f^{ - \n} \lb G_0 (r) + \o G_1 (r)  + \o^2 G_2 (r) \rb    + {\cal O} (\o^3)  ,
\ee 
where $\n = i \fr{\o}{4 r_h^2}$. In this case, 
$\o^2 G_2(r) $ is important to determine the DC conductivity unlike the previous case.
The solutions, $G_0 (r) $ and $G_1 (r) $, satisfying the regularity and vanishing condition at the horizon are
\bea		\la{res:psolg2}
G_0 (r) &=& c_1 , \nn
G_1 (r) &=& - \lb \fr{i}{2 r_h^4}  r^2 - \fr{\pi + 2 i (1 - \log 2 )}{4 r_h^2} 
+ \fr{i }{r_h^2} \log r - \fr{i }{2 r_h^2} \log \ls r^2 + r_h^2 \rs \rb c_1 .
\eea
After inserting these two solutions into \eq{eq:impurityeq}, we can find the analytic form
of $G_2 (r)$ which has the following expansion near the horizon
\bea
G_2 (r) &=& \fr{12 c_6 r_h^6 - c_1 (6 \log 2 - 10 - 3 \pi i  )}{48 rh^4} \log (r - r_h) 
+ c_5 
 + \fr{r_h^2  (2 + \pi i - \log r_h)}{4} c_6 \nn
&&+ \fr{5 \pi^2 + 8 (9 - 10 \log 2 + 3 ( \log 2 )^2) + 
 12 \pi i  (3 + 2 \log 2 ) }{96 r_h^4} c_1 \nn
&& + \fr{(-20 + 42 \pi i + 60 \log 2 ) \log r_h + 
 48 (\log r_h )^2}{96 r_h^4} c_1  + {\cal O} (r-r_h) .
\eea
Again, imposing the regularity and vanishing condition at the horizon, $c_5$ and $c_6$ are fixed 
to be
\bea   \la{res:psol2g2}
c_5 &=& - \lb \fr{32 + 4\pi i + 11 \pi^2 - 56 \log 2  
+ 36 \pi i \log 2 + 24 (\log 2)^2}{96 r_h^4} 
+ \fr{48 (\pi i +\log 2) \log r_h}{96 r_h^4}  + \fr{(\log r_h )^2}{2 r_h^4}  \rb c_1  , \nn
c_6 &=& \fr{6 \log 2 - 10 - 3 \pi i }{12 r_h^6} c_1 .
\eea

Before calculating the conductivity, it is worth to note that 
for $\g \ge 1$ the second terms in \eq{res:asymtsoldu1} is more dominant
when determining the asymptotic behavior of impurity.
This implies that the previous Dirichlet boundary condition in \eq{bc:diffu1so} can not 
fix $b_1$, so we need to modify the asymptotic boundary condition.
A natural choice is choosing the second coefficient $b_2$ 
as a source rather than the first and then fixing it by an appropriate boundary condition. 
Following this strategy, the appropriate asymptotic boundary condition for $\g \ge 1$ 
should be \cite{Son:2002sd}
\be
b_0 = \lim_{r_0 \to \infty} \fr{b_i (r_0)}{{r_0}^{2 \g - 2}}  .
\ee
Especially, for $\g =2$ the boundary condition reduces to
\be			\la{bc:g2}
b_0 = \lim_{r \to \infty} \fr{b_i (r_0)}{ r_0^2 }  .
\ee 
Using \eq{res:psolg2} and \eq{res:psol2g2} together with the exact 
solution for $G_2 (r)$, we can easily find the asymptotic expansion of $b_i (r)$ up to $\o^2$
and comparing it with the boundary condition in \eq{bc:g2} determines $c_1$ in terms of $b_0$
\be
c_1 = \fr{24 i r_h^4 }{\o \lb 12 r_h^2  + i \o \ls 6 \log 2 - 10 - 3 \pi i   \rs \rb}   b_0  .
\ee
At first glance, it looks extraordinary because $c_1$ is proportional to $\o^{-1}$. However,
near the horizon it still becomes a solution whose normalization constant is 
proportional to $\o^{-1}$. 

Since the boundary action for $\g= 2$ is given by
\be
S_{B}  =  - \frac{1}{16 \pi G} \int d^3 x  \ \ r_0 
\ b_{0} \ {b_i}' ,
\ee
only $b_i' \sim r_0^{-1}$ can provide the finite contribution to the boundary action. 
If $\fr{\pa}{\pa r} \ls f^{- \n} G_0  \rs$ or 
$\fr{\pa}{\pa r} \ls f^{- \n} \o G_1 \rs$ contains such a term, the DC
conductivity diverges with $\o^{-2}$ or $\o^{-1}$ respectively because $c_1 \sim 1/ \o$.
This fact says that the finite contribution to the DC conductivity is determined 
not by  $\o G_1$ but by $\o^2 G_2$. 
In the asymptotic region ($r_0 \to \infty$), the expansion of $b_i '$ has the following form 
\be
b_i ' = \# r_0 + \fr{\#}{r_0^3} + {\cal O} \ls \fr{1}{r_0^5} \rs ,
\ee
where $\#$ implies a certain number. This result shows that $b_i'$ has no term 
proportional to $r_0^{-1}$ so that the finite part of the resulting boundary action 
becomes zero. 
Consequently, the DC conductivity carried by impurity for $\g = 2$ vanishes
\be 		\la{res:dccong2}
\s_{DC} = 0 \quad \text{for} \ \g=2 .
\ee 
 
Before concluding this section, there is an important remark.
The result in \eq{res:dccong2} is totally different from that of the membrane paradigm.
In the membrane paradigm, the DC conductivity especially for impurity
can be represented only by the horizon quantities. 
The resulting form for $z=2$ is \cite{Park:2012cu,Kovtun:2003wp,Iqbal:2008by}
\be			\la{res:dccondb}
\s_{DC} = \left. \fr{e^{\g \ph}}{16 \pi G} \sqrt{\fr{g}{g_{tt} g_{rr}}} g^{ii} \right|_{r_h} 
= \fr{1}{16 \pi^{\g+1} G} \ \fr{1}{T^{\g}} .
\ee 
This is exactly the form obtained in the previous section for $\g<1$. However, it is not consistent
with the result for $\g=2$. The reason is that in the Kubo formula the asymptotic boundary condition 
is modified for $\g \ge 1$  while the membrane paradigm does not care about the asymptotic
behavior of the solution. Because of that, the membrane paradigm is consistent with the Kubo formula 
only for $\g<1$. 
Following the spirit of the AdS/CFT correspondence, physical quantities like the correlation
functions of the dual field theory should be holographically governed by bulk field 
fluctuations near the asymptotic region. 
In this sense, the Kubo formula results seems to be more fundamental.

\subsection{DC conductivity carried by Lifshitz matter}

Now, consider Lifshitz matter instead of impurity.  
Generally, if there exists a background gauge field, the transverse mode of its fluctuation
should be coupled to the shear mode of the metric fluctuations
\cite{Ge:2008ak}. 
Therefore, in order to study the linear response of such an U(1) gauge field, we should also 
consider the metric shear mode. 
The equations governing the transverse and shear modes can be derived from \eq{act:flucta}.
After the Fourier mode expansion 
\bea
g^i_t (t,r) &=& \int \fr{d \o}{2 \pi} e^{- i \o t} g^i_t (\o,r) ,  \nn
a_i (t,r) &=& \int \fr{d \o}{2 \pi} e^{- i \o t} a_i (\o,r) ,
\eea 
the governing equations for shear modes reduce to
\bea
0 &=& {g^i_t}^{\pr}  +  \frac{q }{r^{5-z}}  a_i  , 
\la{eq:constraint} \\
0 &=& {g^i_t}^{\pr\pr} + \frac{(5-z)  }{r} {g^i_t}^{\pr}  + \frac{q }{r^{5-z}} a_i^{\pr} , 
\la{eq:gxy}
\eea
where $g^i_t$ and $a_i$ imply $g^i_t (\o,r) $ and $a_i (\o,r)$. 
The first equation \eq{eq:constraint} is a constraint which automatically satisfies the
second equation \eq{eq:gxy}.
The equation governing the transverse modes leads to
\be	 \la{eq:a}
0= a_i^{\pr \pr} +\frac{ r f^{\pr} + (z-3) f  }{r  f} a_i^{\pr}
+\frac{q r^{3-z} }{f } {g^i_t}^{\pr} +\frac{w^2}{r^{2 + 2 z } f ^2} a_i .
\ee
Inserting the constraint into \eq{eq:a}, the decoupled differential
equation of the transverse mode becomes
\be			\la{eq:decoupledAx}
0= a_i^{\pr \pr} +\frac{ r f^{\pr} + (z-3) f  }{r  f} a_i^{\pr}
 + \ls \frac{w^2}{r^{2 z + 2} f ^2}  - \frac{q^2 }{r^2 f }  \rs  a_i .
\ee

At the horizon, due to vanishing of $f$, $a_i$ should have the following two independent solutions up to 
an overall normalization constant
\be			\la{sol:hor}
a_i \sim f^{\pm i \fr{\o}{4 r_h^2} } ,
\ee
where the plus or minus sign again implies the outgoing or incoming mode.
In the hydrodynamic limit, after taking only the incoming part, the near horizon solution of $a_i$ can be expanded
into 
\be		\la{ans:pertsol}
a_i = f^{- i \fr{\o}{4 r_h^2} } \ \lb G_0 (r) + \o \ G_1 (r)  \rb 
+ {\cal O} (\o^2) ,
\ee
where $G_0 (r)$ and $G_1 (r)$ should be regular functions at the black brane
horizon. 
After substituting the perturbative expansion form to \eq{eq:decoupledAx}, 
one can solve it order by order. For $z=2$,
$G_0$ has the following exact solution at $\o^0$ order
\be
G_0 = c_1 \ls r^4 + r_h^4 \rs - \fr{c_2 \ r^2}{8 r_h^4} 
+ \fr{c_2 \ (r^4 + r_h^4)}{8 r_h^6} \ {\rm arctanh} \ls \fr{r^2}{r_h^2} \rs,
\ee
where $c_1$ and $c_2$ are two integration constants. Imposing the regularity condition
at the horizon, $c_2 = 0$  because ${\rm arctanh} \ls
\fr{r^2}{r_h^2} \rs$ diverges. Consequently, 
\be
G_0 = c_1 \ls r^4 + r_h^4 \rs .
\ee 
Using this result, at next order of $\o$ the solution of $G_1$ is given by
\bea
G_1 &=& c_3  (r^4 + r_h^4) - \fr{c_4 r^2}{8 r_h^4}  - \fr{i c_1  (r^4 + r_h^4) \log r}{r_h^2}
+ \fr{  (r^4 + r_h^4) ( c_4 + i 4 c_1 r_h^4) \ \log (r^2 + r_h^2)}{16 r_h^6} \nn
&& - \fr{  (r^4 + r_h^4) (c_4 - i 4 c_1 r_h^4)  \log (r^2 - r_h^2)}{16 r_h^6} .
\eea
Since the last term diverges at the horizon, the regularity condition determines
\be
c_4 = i 4 c_1 r_h^4.
\ee 
In addition, the vanishing condition fixes the rest integration
constant as 
\be
c_3 =  - \fr{i c_1 \ (2 \log 2  -1 )}{4 r_h^2} .
\ee

Now, let us consider the asymptotic behavior of the solution. Unlike the impurity case, 
the asymptotic behavior of $a_i$ is totally different from that of $b_i$
due to the nontrivial mixing with the shear mode. From \eq{eq:decoupledAx},
the asymptotic behavior for $z=2$ is governed by
\be			
0= a_i^{\pr \pr} - \frac{1}{r} a_i^{\pr} - \frac{8 }{r^2}   a_i ,
\ee
where the last term is originated from the shear mode.  
At the asymptotic boundary, $a_i$ has the following perturbative solution 
\bea			\la{bc:lifgauge}
a_i =  a_1  r^4 +  \fr{ a_2}{r^2}   ,
\eea
where $a_1$ is a constant to be determined by the asymptotic boundary condition. 
One can identify the coefficients of the non-normalizable and normalizable modes, $a_1$ and $a_2$, with the source and expectation value of the dual operator respectively.
In order to fix the boundary value of $a_i$, we impose the following
boundary condition at the asymptotic boundary 
\be			\la{bdc:Ax}
a_0 \equiv \lim_{r \to \infty} \frac{a_i (r)}{r^4}  .
\ee 
Then, the integration constant $c_1$ is fixed in terms of the boundary value $a_0$ to be
\be
c_1 = \fr{4 \ r_h^2 \ a_0}{ 4 r_h^2 - i  \o (2 \log 2 - 1)  }    .
\ee

From the action \eq{act:flucta}, the boundary term of the Lifshitz matter becomes
\bea \la{bact:gauge}
S_B &=& - \fr{1}{16 \pi G} \int d^3 x \  \sqrt{-g} \ e^{\l \ph} g^{rr} g^{ii}  a_i \pa_r a_i 
\nn
&=& - \fr{1}{16 \pi G} \int d^3 x  \ r_0^3  \ a_0 \ \pa_r a_i ,
\eea
where one can see that the finite part of the retarded Green function comes from 
$\pa_r a_i \sim r_0^{- 3}$. When ignoring the divergent parts corresponding
to the contact terms, the DC conductivity carried by the Lifshitz matter
leads to
\be			\la{res:kuboa}
\s_{DC} = \fr{\pi}{12  G}  \  T^2.
\ee
This result shows that the DC conductivity carried by a Lifshitz matter is
totally different from that carried by impurity for $\g=2$.

Impurity in semiconductor dramatically changes the electric property of matter from an insulator
at low temperature to a conductor at high temperature. 
In many condensed matter systems like a semiconductor, impurity plays a crucial role in 
physics so that it is important to understand the effect of such impurity. In this paper, we holographically
realize impurity by turning on a different bulk vector fluctuation in the Lifshitz black brane geometry. 
In the dual theory point of view, it corresponds to inserting impurity into the Lifshitz medium.
Depending on the kind of impurity parameterized by $\g$, it can dramatically change the
electric property of the Lifshitz matter. 
For instance, if impurity is characterized by $\g<-2$, the DC conductivity
of the Lifshitz matter is not affected by impurity at low temperature. However, at high temperature
the DC conductivity of impurity is dominant so that the electric property of the Lifshitz matter with
impurity can show totally different behavior from the pure Lifshitz matter.


\section{Discussion}

We have investigated thermodynamics and hydrodynamics of the non-relativistic 
Lifshitz medium with two types of charge carriers. To do so, we considered a Lifshitz black brane
solutions of the Einstein-Maxwell-dilaton gravity. There are several models describing the same
asymptotic Lifshitz geometry. Although all models give rise to the same thermodynamic relation
and the equation of state parameter, the details of the thermodynamic quantities are different
due to the different interior metric. In this paper, we clarified the scaling behaviors of all fields
and thermodynamic quantities of the Lifshitz black brane geometry derived from the Einstein-Maxwell-dilaton
gravity. 

After that, we have studied the holographic responses of two types of vector fluctuations  
which describe the DC conductivities carried by two different charge carriers, impurity and 
the Lifshitz matter, in the non-relativistic Lifshitz medium. 
In this case, to realize impurity on the dual
gravity we have turned on another vector fluctuation in the Lifshitz black brane geometry.
For the DC conductivity carried by impurity, there are two different methods, the Kubo formula and
membrane paradigm. 
n the non-relativistic Lifshitz medium, these two formula give the same result for $\g<1$ 
but we found that there exists a marked discrepancy for $\g \ge 1$. 
For $\g \ge 1$,  one should change the asymptotic boundary condition to find the DC conductivity. This fact
can be consistently imposed in the Kubo formula, while the membrane paradigm, because it is described by
only information at the horizon, can not know this fact. In that sense, the Kubo formula looks more
fundamental.

For the DC conductivity carried by the Lifshitz matter, its dual vector fluctuation is coupled to the 
metric fluctuation through the background gauge field. The coupling with the metric fluctuation
leads to a nontrivial DC conductivity proportional to the square of temperature for $z=2$. 
Like an example of semiconductor, we found that impurity with $\g<-2$ 
can crucially modify the electric property of the non-relativistic
Lifshitz matter at high temperature.

\vspace{1cm}

{\bf Acknowledgement}

This work has been supported by the WCU grant no. R32-10130 and the Research fund no. 1-2008-2935-001-2
by Ewha Womans University.  
C. Park was also supported by Basic Science Research Program through the National Research Foundation of 
Korea(NRF) funded by the Ministry of Education (NRF-2013R1A1A2A10057490). 

\vspace{1cm}


\end{document}